\documentclass[twocolumn,superscriptaddress,preprintnumbers,nofootinbib]{revtex4}
\usepackage{amsmath,amssymb}
\usepackage{graphicx}
\usepackage{epsfig,color}
\usepackage{bm}
\usepackage{verbatim}

\begin{document}

\title{Hidden-charm and bottom tetra- and pentaquarks with strangeness in the hadro-quarkonium and compact tetraquark models}

\author{J. Ferretti}\email[]{jacopo.j.ferretti@jyu.fi}
\affiliation{Center for Theoretical Physics, Sloane Physics Laboratory, Yale University, New Haven, Connecticut 06520-8120, USA}
\affiliation{Department of Physics, University of Jyv\"askyl\"a, P.O. Box 35 (YFL), 40014 Jyv\"askyl\"a, Finland}
\author{E. Santopinto}\email[]{elena.santopinto@ge.infn.it}
\affiliation{Istituto Nazionale di Fisica Nucleare (INFN), Sezione di Genova, Via Dodecaneso 33, 16146 Genova, Italy}

\begin{abstract}
In two recent papers, we used the hadro-quarkonium model to study the properties of hidden-charm and bottom tetraquarks and pentaquarks. 
Here, we extend the previous results and calculate the masses of heavy-quarkonium-kaon/hyperon systems.
We also compute the spectrum of hidden-charm and bottom tetraquarks with strangeness in the compact tetraquark (diquark-antidiquark) model.
If heavy-light exotic systems with non-null strangeness content were to be observed experimentally, it might be possible to distinguish among the large variety of available theoretical pictures for tetra- and pentaquark states and, possibly, rule out those which are not compatible with the data.
\end{abstract}

\maketitle

\section{Introduction}
Multiquark states are baryons/mesons which cannot be described in terms of $qqq$/$q \bar q$ degrees of freedom only.
They include $XYZ$ suspected tetraquarks, like the $X(3872)$ [now $\chi_{\rm c1}(3872)$] \cite{Choi:2003ue,Acosta:2003zx,Abazov:2004kp} and $X(4274)$ [also known as $\chi_{\rm c1}(4274)$] \cite{Aaltonen:2011at,Aaij:2016iza}, and pentaquark states.
The latter were recently discovered by LHCb in $\Lambda_{\rm b} \rightarrow J/\psi \Lambda^*$ and $\Lambda_{\rm b} \rightarrow P_{\rm c}^+ K^- \rightarrow (J/\psi p) K^-$ decays \cite{Aaij:2015tga,Aaij:2019vzc}.
The structure of $XYZ$ tetraquarks and $P_{\rm c}$ pentaquarks is still unclear.
This is why there are several alternative models to explain their properties. For a review, see Refs. \cite{Chen:2016qju,Ali:2017jda,Olsen:2017bmm,Guo:2017jvc}.
To distinguish among the different pictures (molecular model, diquark model, unquenched quark model, ...) one should compare their theoretical predictions for the spectrum, decay amplitudes, production cross-sections, and so on, with the experimental data.

A clean way to discriminate among the previous theoretical interpretations for suspected $XYZ$ tetraquarks was suggested in Ref. \cite{Voloshin:2019}.
There, Voloshin pointed out that if $Z_{\rm c}$ resonances exist then, because of the SU(3)$_{\rm f}$ symmetry, one may also expect the emergence of their strange partners, $Z_{\rm cs}$ \cite{Voloshin:2019}.
The author also argued that the one-pion-exchange interaction of the meson-meson molecular model is impossible between strange and nonstrange heavy mesons, like $B$ and $B_{\rm s}$ \cite{Voloshin:2019}.
Hidden-charm and bottom mesons with strangeness are also forbidden in the context of the Unquenched Quark Model (UQM) formalism. Indeed, one cannot dress heavy quarkonium $Q \bar Q$ states with $Q \bar s - n \bar Q$ or $Q \bar n - s \bar Q$ higher Fock components (where $n = u$ or $d$) by creating a light $n \bar n$ or $s \bar s$ pair with vacuum quantum numbers.
Therefore, hidden-charm and bottom tetraquark states with non-null strangeness content cannot take place neither in the UQM \cite{Heikkila:1983wd,Pennington:2007xr,Danilkin:2010cc,Ortega:2012rs,Ferretti:2013faa,Ferretti:2014xqa,Ferretti:2013vua,Achasov:2015oia,Kang:2016jxw,Lu:2016mbb,Ferretti:2018tco} nor in the molecular model \cite{Tornqvist:1993ng,Hanhart:2007yq,Baru:2011rs,Valderrama:2012jv,Aceti:2012cb,Guo:2013sya} interpretations.
\begin{figure}[htbp]
\begin{minipage}{9pc}
\includegraphics[width=8.5pc]{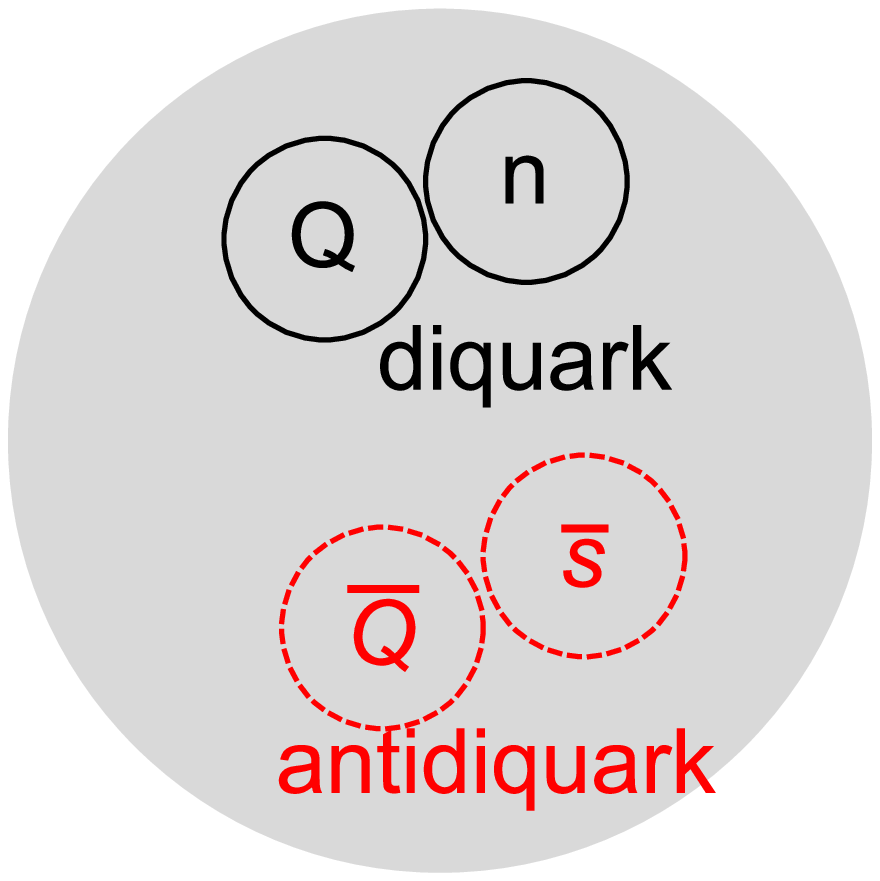}
\end{minipage}
\hspace{1pc}
\begin{minipage}{9pc}
\includegraphics[width=8.5pc]{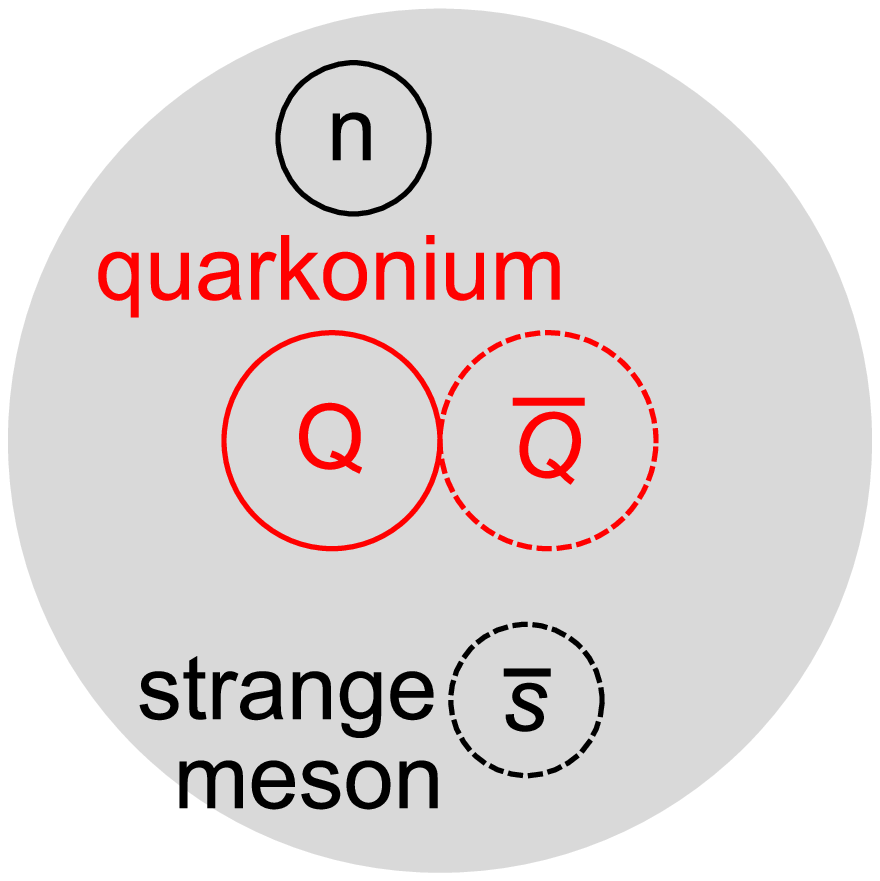}
\end{minipage}
\caption{Schematic representation of heavy-light hadro-quarkonium (right) and compact tetraquark (left) states.}
\label{fig:4-quarks}
\end{figure} 
On the contrary, these exotic configurations are expected (if above threshold) both in the compact tetraquark \cite{Jaffe:1976ih,SilvestreBrac:1993ss,Brink:1998as,Maiani:2004vq,Barnea:2006sd,Santopinto:2006my,Ebert:2008wm,Deng:2014gqa,Zhao:2014qva,Lu:2016cwr,Anwar:2017toa,Anwar:2018sol,Esposito:2018cwh,Bedolla:2019zwg,Yang:2019itm} and hadro-quarkonium \cite{Eides:2015dtr,Perevalova:2016dln,Alberti:2016dru,Anwar:2018bpu,Dubynskiy:2008mq,Guo:2008zg,Voloshin:2013dpa,Wang:2013kra,Brambilla:2015rqa,Ferretti:2018kzy,Panteleeva:2018ijz,Voloshin:2019} models.
See Fig. \ref{fig:4-quarks}.
In light of this, the experimental observation of $XYZ$ states with non-null strangeness content would make it possible to rule out a few possible theoretical interpretations for tetraquarks.
Voloshin did not compute the spectrum of $Z_{\rm cs}$ states, but only discussed phenomenological indications for the emergence of those states \cite{Voloshin:2019}.
The study of their spectrum and that of their pentaquark counterparts is thus the subject of the present manuscript.

Here, we extend the hadro-quarkonium model findings of Refs. \cite{Ferretti:2018kzy,Anwar:2018bpu} and calculate the spectrum of hidden-charm and bottom tetraquarks and pentaquarks with strangeness. 
The hadro-quarkonium picture was developed to explain the experimental observation of heavy-light tetraquark candidates characterized by peculiar properties \cite{Dubynskiy:2008mq,Voloshin:2007dx}. Firstly, these exotics are supposed not to be particularly close to a specific heavy-light meson-meson threshold, unlike $D^0 \bar D^{*0}$ in the $X(3872)$ case. Secondly, such states may decay into  heavy quarkonia plus one or more light mesons, like $\eta_{\rm c} + \eta$.
Even though it was meant for the description of tetraquarks, the hadro-quarkonium model can be easily extended to the baryon sector to study pentaquarks \cite{Anwar:2018bpu,Eides:2015dtr}.

We also compute the masses of heavy-light tetraquarks with non-null strangeness content in the compact tetraquark model of Refs. \cite{Anwar:2017toa,Anwar:2018sol,Bedolla:2019zwg}.
In the compact tetraquark model, heavy-light $qQ \bar q \bar Q$ states are modeled as the bound states of a diquark, $qQ$, antidiquark, $\bar q \bar Q$, pair. The diquark constituents are treated as inert against internal spatial excitations. 
Their binding is the consequence of one-gluon-exchange forces and their relative dynamics can be described in terms of a relative coordinate $\bf r_{\rm rel}$.
The calculation of the spectrum of compact pentaquark configurations in the diquark model is more difficult than that of compact tetraquarks because one has to deal with a three-body problem instead of a two-body one; moreover, one also has to consider both diquark-diquark and diquark-antiquark interactions. This is why here we do not provide results for compact (diquark-diquark-antiquark) pentaquarks, which will be the subject of a subsequent paper.

Our predictions for strange hidden-charm and bottom tetraquarks and, especially, those for $P_{\rm c}$ and $P_{\rm b}$ pentaquarks with non-null strangeness content may soon be tested by LHCb.

\section{Hadro-quarkonium model}
\label{Hadro-quarkonium Hamiltonian}
The possible existence of binding mechanisms of charmonium states in light-quark matter was discussed long ago \cite{Brodsky:1989jd,Kaidalov:1992hd,Sibirtsev:2005ex} in terms of the interaction of charmonium inside nuclei. 
The idea of hadro-charmonium (hadro-quarkonium) bound states resembles the previous one.

Hadro-quarkonia are heavy-light tetra- or pentaquark configurations, where a compact $Q \bar Q$ state (with $Q = c$ or $b$), labelled as $\psi$ in the following, is embedded in light hadronic matter, $\mathcal H = $ $qqq$ or $q \bar q$ (where $q = u, d$ or $s$) \cite{Eides:2015dtr,Perevalova:2016dln,Alberti:2016dru,Anwar:2018bpu,Dubynskiy:2008mq,Guo:2008zg,Voloshin:2013dpa,Wang:2013kra,Brambilla:2015rqa,Ferretti:2018kzy,Panteleeva:2018ijz,Voloshin:2019}.
The heavy and light constituents, $\psi$ and $\mathcal H$, develop an attractive force, which is the result of multiple-gluon exchange between them.
Such interaction, $H_{\rm eff}$, can be written in terms of the multipole expansion in QCD \cite{QCDME}. 
In particular, if one considers as leading term the $E1$ interaction with chromo-electric fields ${\bf E}$ and ${\bf E}'$ \cite{Dubynskiy:2008mq,Kaidalov:1992hd}, one gets the effective Hamiltonian
\begin{equation}
	\label{eqn:Heff}
	H_{\rm eff} = - \frac{1}{2} \alpha_{\psi\psi'} {\bf E} \cdot {\bf E}'  \mbox{ },
\end{equation}
where $\alpha_{\psi\psi'}$ is the so-called heavy quarkonium chromo-electric polarizability.
By making use of additional approximations, $H_{\rm eff}$ can be further reduced to a simple square-well potential \cite{Dubynskiy:2008mq,Ferretti:2018kzy,Anwar:2018bpu},
\begin{equation}
	\label{eqn:Vhc}
	V_{\rm hq}(r) = \left\{ \begin{array}{ccc} -\frac{2\pi\alpha_{\psi\psi}M_{\mathcal H}}{3R_{\mathcal H}^3} & \mbox{for} & 
	r < R_{\mathcal H} \\ 0   & \mbox{for} & r > R_{\mathcal H} \end{array}  \right.  \mbox{ },
\end{equation}
where $R_{\mathcal H} = R_{\mathcal B}$ or $R_{\mathcal M}$ is the light baryon/meson radius. 
Eq. (\ref{eqn:Vhc}) can be plugged into a Schr\"odinger equation and solved for light hadron-heavy quarkonium systems.
\begin{figure}[htbp] 
\centering 
\includegraphics[width=5cm]{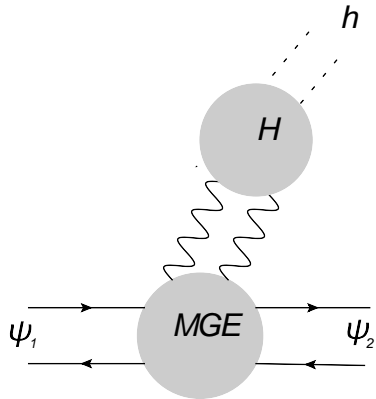}
\caption{Hidden-flavor transition $\psi_1 \rightarrow \psi_2 + h$ in the QCD multipole expansion. Here, $\psi_1$ and $\psi_2$ are the initial and final charmonium states, $h$ light hadron(s). The two vertices are those of the multipole gluon emission, $MGE$, and hadronization, $H$. Picture from Ref. \cite{Ferretti:2018kzy}; Elsevier Copyright.}
\label{fig:gluon-trans}
\end{figure}

There are four quantities to be given as input in the calculation.
They are the masses $M_\psi$ and $M_{\mathcal H}$, the radius $R_{\mathcal H}$, and the diagonal chromo-electric polarizability, $\alpha_{\psi\psi}$. See Table \ref{tab:Hq-Model-parameters}.
\begin{table}[htbp]
\centering
\begin{ruledtabular}
\begin{tabular}{cccc}
Parameter  & Value   & Parameter  & Value \\
\hline
$\alpha_{\psi \psi}(1P)_{c \bar c}$ & 11 GeV$^{-3}$ &  $\alpha_{\psi \psi}(2S)_{c \bar c}$ & 18 GeV$^{-3}$ \\
$\alpha_{\psi \psi}^{(1)}(1P)_{b \bar b}$ & 14 GeV$^{-3}$ & $\alpha_{\psi \psi}^{(1)}(2S)_{b \bar b}$ & 23 GeV$^{-3}$ \\
$\alpha_{\psi \psi}^{(2)}(1P)_{b \bar b}$ & 21 GeV$^{-3}$ & $\alpha_{\psi \psi}^{(2)}(2S)_{b \bar b}$ & 33 GeV$^{-3}$ \\
$R_{\Sigma}$ & 0.863 fm & $R_{\Xi}$ & 0.841 fm \\
$R_K$ & 0.560 fm & $R_{K^*}$ & 0.729 fm \\
\end{tabular}
\end{ruledtabular}
\caption{Hadro-quarkonium model. Input values and parameters.}
\label{tab:Hq-Model-parameters}
\end{table}
The values of $M_\psi$ and $M_{\mathcal H}$ are extracted from the PDG \cite{Tanabashi:2018oca}.

In principle, non-diagonal quarkonium chromo-electric polarizabilities, $\alpha_{\psi\psi'}$, can be fitted to the data by considering $\psi \rightarrow \psi' + h$ hadronic transitions \cite{Voloshin:2007dx,Chen:2019gty}; see Fig. \ref{fig:gluon-trans}.
However, no experimental information can be used to estimate the $\alpha_{\psi\psi}$'s.
Therefore, the diagonal chromo-electric polarizabilities, $\alpha_{\psi \psi}(n \ell)$, where $n$ and $\ell$ are the radial quantum number and orbital angular momentum of $\psi$, respectively, have to be extracted from the phenomenology.
In the case of charmonia, we consider \cite{Anwar:2018bpu}: $\alpha_{\psi \psi}(1P)_{c \bar c} = 11 \mbox{ GeV}^{-3}$ and $\alpha_{\psi \psi}(2S)_{c \bar c} = 18 \mbox{ GeV}^{-3}$.
In the case of bottomonia, we make use of two sets of values for the chromo-electric polarizabilities. They are \cite{Anwar:2018bpu}: 
$\alpha_{\psi \psi}^{(1)}(1P)_{b \bar b} = 14 \mbox{ GeV}^{-3}$ and $\alpha_{\psi \psi}^{(1)}(2S)_{b \bar b} = 23 \mbox{ GeV}^{-3}$; $\alpha_{\psi \psi}^{(2)}(1P)_{b \bar b} = 21 \mbox{ GeV}^{-3}$ and $\alpha_{\psi \psi}^{(2)}(2S)_{b \bar b} = 33 \mbox{ GeV}^{-3}$.
We also need the strange mesons' and hyperons' radii.
While for the kaon we can use the well-established value of the $K^\pm$ charge radius reported on the PDG \cite{Tanabashi:2018oca}, $R_K = 0.560\pm0.031$ fm, in the $\Sigma$, $\Xi$ and $K^*$ cases the situation is different\footnote{The values of the proton and kaon radii reported by the PDG \cite{Tanabashi:2018oca} can be regarded as reliable, because they are the result of the average over several measurements. On the contrary, the value of the $\Sigma^-$ radius from the PDG is the outcome of a single experiment; moreover, there is no available data for the charge radius of the $\Sigma^+$. This is why here we do not extract $R_{\Sigma}$ from the PDG.}.
Indeed, due to the lack of well-established experimental data, we are forced to extract $R_{\Sigma}$, $R_{\Xi}$ and $R_{K^*}$ from phenomenological estimates. For example, see Refs. \cite{Ledwig:2006gw,Kubis:1999xb,Buchmann:2002et,Carrillo-Serrano:2016igi,Godfrey:1985xj}.
Following Ref. \cite{Ledwig:2006gw}, we have: $R_{\Sigma} = \frac{1}{2} \left(R_{\Sigma^+} + R_{\Sigma^-}\right) = 0.863$ fm; $R_{\Xi} = 0.841$ fm.
The $K^*(892)$'s radius is calculated in the relativized quark model for mesons of Ref. \cite{Godfrey:1985xj}: $R_{K^*} = 0.729$ fm.

Finally, the hadro-quarkonium quantum numbers are obtained by combining those of the hadrons $\mathcal \psi$ and $\mathcal H$, 
\begin{equation}
\label{quantNumbers}
	\left| \Phi_{\rm hq} \right\rangle = \left| (L_\psi, S_\psi) J_\psi; (L_{\mathcal H}, S_{\mathcal H}) J_{\mathcal H}; 
	(J_{\rm hq}, \ell_{\rm hq}) J_{\rm tot}^P \right\rangle  \mbox{ }.
\end{equation}
Here, ${\bf J}_{\rm hq} = {\bf J}_\psi + {\bf J}_{\mathcal H}$, the hadro-quarkonium parity is $P = (-1)^{\ell_{\rm hq}}$ $P_\psi P_{\mathcal H}$, and $\ell_{\rm hq}$ is the relative angular momentum between $\psi$ and $\mathcal H$. 
From now on, unless explicitly indicated, we assume that $\ell_{\rm hq} = 0$.
\begin{table*}[htbp]
\begin{ruledtabular}
\begin{tabular}{cccccccccc}
Composition & & Quark content & & $\alpha_{\psi \psi}(n\ell)$ [GeV$^{-3}$] & & $J_{\rm tot}^P$ & & Mass (Binding) [MeV] \\
\hline
$\chi_{\rm c0}(1P) \otimes \Sigma$ & & $nnsc \bar c$ & & 11 & & $\frac{1}{2}^+$ & & 4440 ($-166$) \\
$\eta_{\rm c}(2S) \otimes \Sigma$ & & $nnsc \bar c$ & & 18 & & $\frac{1}{2}^-$ & & 4474 ($-355$)  \\
$\psi(2S) \otimes \Sigma$ & & $nnsc \bar c$ & & 18 & & $\frac{1}{2}^-$ or $\frac{3}{2}^-$ & & 4522 ($-355$)  \\
$\chi_{\rm c1}(1P) \otimes \Sigma$ & & $nnsc \bar c$ & & 11 & & $\frac{1}{2}^+$ or $\frac{3}{2}^+$ & & 4535 ($-166$) \\
$h_{\rm c}(1P) \otimes \Sigma$ & & $nnsc \bar c$ & & 11 & & $\frac{1}{2}^+$ or $\frac{3}{2}^+$ & & 4550 ($-167$) \\
$\chi_{\rm c2}(1P) \otimes \Sigma$ & & $nnsc \bar c$ & & 11 & & $\frac{3}{2}^+$ or $\frac{5}{2}^+$ & & 4580 ($-167$) \\
$\left[\eta_{\rm c}(2S) \otimes \Sigma\right]_{\ell_{\rm hq} = 1}$ & & $nnsc \bar c$ & & 18 & & $\frac{1}{2}^+$ or $\frac{3}{2}^+$ & & 4653 ($-175$)  \\
$\left[\psi(2S) \otimes \Sigma\right]_{\ell_{\rm hq} = 1}$ & & $nnsc \bar c$ & & 18 & & $\frac{1}{2}^+$, $\frac{3}{2}^+$ or $\frac{5}{2}^+$ & & 4701 ($-176$) \\
\hline
$\eta_{\rm c}(2S) \otimes \Xi$ & & $nssc \bar c$ & & 18 & & $\frac{1}{2}^-$ & & 4500 ($-459$); 4955 ($-5$) \\
$\chi_{\rm c0}(1P) \otimes \Xi$ & & $nssc \bar c$ & & 11 & & $\frac{1}{2}^+$ & & 4510 ($-226$) \\
$\psi(2S) \otimes \Xi$              & & $nssc \bar c$ & & 18 & & $\frac{1}{2}^-$ or $\frac{3}{2}^-$ & & 4548 ($-460$); 5002 ($-5$) \\
$\chi_{\rm c1}(1P) \otimes \Xi$ & & $nssc \bar c$ & & 11 & & $\frac{1}{2}^+$ or $\frac{3}{2}^+$ & & 4605 ($-227$) \\
$h_{\rm c}(1P) \otimes \Xi$ & & $nssc \bar c$ & & 11 & & $\frac{1}{2}^+$ or $\frac{3}{2}^+$ & & 4620 ($-227$) \\
$\chi_{\rm c2}(1P) \otimes \Xi$ & & $nssc \bar c$ & & 11 & & $\frac{3}{2}^+$ or $\frac{5}{2}^+$ & & 4650 ($-228$) \\
$\left[\eta_{\rm c}(2S) \otimes \Xi\right]_{\ell_{\rm hq} = 1}$ & & $nssc \bar c$ & & 18 & & $\frac{1}{2}^+$ or $\frac{3}{2}^+$ & & 4685 ($-274$)  \\
$\left[\psi(2S) \otimes \Xi\right]_{\ell_{\rm hq} = 1}$ & & $nssc \bar c$ & & 18 & & $\frac{1}{2}^+$, $\frac{3}{2}^+$ or $\frac{5}{2}^+$ & & 4733 ($-275$) \\                                             
\end{tabular}
\end{ruledtabular}
\caption{Hadro-quarkonium model predictions for charmonium-$\Sigma$ and $\Xi$ bound states. The pentaquark binding energies and masses (5th column) are calculated with the values of the chromo-electric polarizabilities $\alpha_{\psi \psi}(n\ell)$ (3rd column). Here, $n = u$ or $d$. The bound states are $S$-wave configurations (i.e. $\ell_{\rm hq} = 0$), except where explicitly indicated. In some cases, the $V_{\rm hq}$ potential well is deep enough to give rise to a heavy-quarkonium$-$baryon bound state and its radial excitation. In this instance, the masses of both the ground-state and excited hadro-quarkonium configurations are reported in the fifth column.}
\label{tab:hadro-quarkonium-spectrum-ccS}
\end{table*}

\begin{table*}[htbp]
\begin{ruledtabular}
\begin{tabular}{cccccccccc}
Composition & & Quark content & & $\alpha_{\psi \psi}(n\ell)$ [GeV$^{-3}$] & & $J_{\rm tot}^P$ & & Mass (Binding) [MeV] \\
\hline
$\eta_{\rm b}(2S) \otimes \Sigma$ & & $nnsb \bar b$ & & 23 & & $\frac{1}{2}^-$ & & 10671 ($-519$); 11118 ($-72$)   \\
$\Upsilon(2S) \otimes \Sigma$ & & $nnsb \bar b$ & & 23 & & $\frac{1}{2}^-$ or $\frac{3}{2}^-$ & & 10695 ($-519$); 11142 ($-72$)  \\
$\chi_{\rm b0}(1P) \otimes \Sigma$ & & $nnsb \bar b$ & & 14 & & $\frac{1}{2}^+$ & & 10784 ($-267$)   \\
$\chi_{\rm b1}(1P) \otimes \Sigma$ & & $nnsb \bar b$ & & 14 & & $\frac{1}{2}^+$ or $\frac{3}{2}^+$ & & 10817 ($-267$)   \\
$h_{\rm b}(1P) \otimes \Sigma$ & & $nnsb \bar b$ & & 14 & & $\frac{1}{2}^+$ or $\frac{3}{2}^+$ & & 10824 ($-267$)  \\
$\chi_{\rm b2}(1P) \otimes \Sigma$ & & $nnsb \bar b$ & & 14 & & $\frac{3}{2}^+$ or $\frac{5}{2}^+$ & & 10836 ($-267$)   \\
$\left[\eta_{\rm b}(2S) \otimes \Sigma\right]_{\ell_{\rm hq} = 1}$ & & $nnsb \bar b$ & & 23 & & $\frac{1}{2}^+$ or $\frac{3}{2}^+$ & & 10840 ($-350$)   \\
$\left[\Upsilon(2S) \otimes \Sigma\right]_{\ell_{\rm hq} = 1}$ & & $nnsb \bar b$ & & 23 & & $\frac{1}{2}^+$, $\frac{3}{2}^+$ or $\frac{5}{2}^+$ & & 10864 ($-350$)  \\
\hline
$\eta_{\rm b}(2S) \otimes \Sigma$ & & $nnsb \bar b$ & & 33 & & $\frac{1}{2}^-$ & & 10383 ($-807$); 10885 ($-305$)   \\
$\Upsilon(2S) \otimes \Sigma$ & & $nnsb \bar b$ & & 33 & & $\frac{1}{2}^-$ or $\frac{3}{2}^-$ & & 10407 ($-808$); 10909 ($-306$)  \\
$\left[\eta_{\rm b}(2S) \otimes \Sigma\right]_{\ell_{\rm hq} = 1}$ & & $nnsb \bar b$ & & 33 & & $\frac{1}{2}^+$ or $\frac{3}{2}^+$ & & 10564 ($-626$); 11175 ($-15$)   \\
$\left[\Upsilon(2S) \otimes \Sigma\right]_{\ell_{\rm hq} = 1}$ & & $nnsb \bar b$ & & 33 & & $\frac{1}{2}^+$, $\frac{3}{2}^+$ or $\frac{5}{2}^+$ & & 10588 ($-626$); 11199 ($-15$)  \\
$\chi_{\rm b0}(1P) \otimes \Sigma$ & & $nnsb \bar b$ & & 21 & & $\frac{1}{2}^+$ & & 10588 ($-462)$; 11016 ($-34$)  \\
$\chi_{\rm b1}(1P) \otimes \Sigma$ & & $nnsb \bar b$ & & 21 & & $\frac{1}{2}^+$ or $\frac{3}{2}^+$ & & 10622 ($-462$); 11049 ($-34$)   \\
$h_{\rm b}(1P) \otimes \Sigma$ & & $nnsb \bar b$ & & 21 & & $\frac{1}{2}^+$ or $\frac{3}{2}^+$ & & 10628 ($-462$); 11056 ($-34$) \\
$\chi_{\rm b2}(1P) \otimes \Sigma$ & & $nnsb \bar b$ & & 21 & & $\frac{3}{2}^+$ or $\frac{5}{2}^+$ & & 10641 ($-462$); 11069 ($-34$)  \\
\hline
$\eta_{\rm b}(2S) \otimes \Xi$ & & $nssb \bar b$ & & 23 & & $\frac{1}{2}^-$ & & 10664 ($-657$); 11126 ($-194$)  \\
$\Upsilon(2S) \otimes \Xi$ & & $nssb \bar b$ & & 23 & & $\frac{1}{2}^-$ or $\frac{3}{2}^-$ & & 10688 ($-657$); 11150 ($-195$)  \\
$\chi_{\rm b0}(1P) \otimes \Xi$ & & $nssb \bar b$ & & 14 & & $\frac{1}{2}^+$ & & 10832 ($-349$)  \\
$\left[\eta_{\rm b}(2S) \otimes \Xi\right]_{\ell_{\rm hq} = 1}$ & & $nssb \bar b$ & & 23 & & $\frac{1}{2}^+$ or $\frac{3}{2}^+$ & & 10833 ($-488$)  \\
$\left[\Upsilon(2S) \otimes \Xi\right]_{\ell_{\rm hq} = 1}$ & & $nssb \bar b$ & & 23 & & $\frac{1}{2}^+$, $\frac{3}{2}^+$ or $\frac{5}{2}^+$ & & 10857 ($-488$) \\
$\chi_{\rm b1}(1P) \otimes \Xi$ & & $nssb \bar b$ & & 14 & & $\frac{1}{2}^+$ or $\frac{3}{2}^+$ & & 10865 ($-349$)  \\
$h_{\rm b}(1P) \otimes \Xi$ & & $nssb \bar b$ & & 14 & & $\frac{1}{2}^+$ or $\frac{3}{2}^+$ & & 10872 ($-349$)  \\
$\chi_{\rm b2}(1P) \otimes \Xi$ & & $nssb \bar b$ & & 14 & & $\frac{3}{2}^+$ or $\frac{5}{2}^+$ & & 10885 ($-349$)   \\
\hline
$\eta_{\rm b}(2S) \otimes \Xi$ & & $nssb \bar b$ & & 33 & & $\frac{1}{2}^-$ & & 10315 ($-1006$); 10818 ($-502$)  \\
$\Upsilon(2S) \otimes \Xi$ & & $nssb \bar b$ & & 33 & & $\frac{1}{2}^-$ or $\frac{3}{2}^-$ & & 10339 ($-1006$); 10842 ($-503$)   \\
$\left[\eta_{\rm b}(2S) \otimes \Xi\right]_{\ell_{\rm hq} = 1}$ & & $nssb \bar b$ & & 33 & & $\frac{1}{2}^+$ or $\frac{3}{2}^+$ & & 10495 ($-826$); 11137 ($-183$)  \\
$\left[\Upsilon(2S) \otimes \Xi\right]_{\ell_{\rm hq} = 1}$ & & $nssb \bar b$ & & 33 & & $\frac{1}{2}^+$, $\frac{3}{2}^+$ or $\frac{5}{2}^+$ & & 10519 ($-826$); 11161 ($-184$)  \\  
$\chi_{\rm b0}(1P) \otimes \Xi$ & & $nssb \bar b$ & & 21 & & $\frac{1}{2}^+$ & & 10593 ($-588$); 11044 ($-138$)   \\
$\chi_{\rm b1}(1P) \otimes \Xi$ & & $nssb \bar b$ & & 21 & & $\frac{1}{2}^+$ or $\frac{3}{2}^+$ & & 10627 ($-588$); 11077 ($-138$) \\  
$h_{\rm b}(1P) \otimes \Xi$ & & $nssb \bar b$ & & 21 & & $\frac{1}{2}^+$ or $\frac{3}{2}^+$ & & 10633 ($-588$); 11083 ($-138$)  \\  
$\chi_{\rm b2}(1P) \otimes \Xi$ & & $nssb \bar b$ & & 21 & & $\frac{3}{2}^+$ or $\frac{5}{2}^+$ & & 10646 ($-588$); 11096 ($-138$) \\                                                
\end{tabular}
\end{ruledtabular}
\caption{As Table \ref{tab:hadro-quarkonium-spectrum-ccS}, but for bottomonium-$\Sigma$ and $\Xi$ bound states.}
\label{tab:hadro-quarkonium-spectrum-ccX}
\end{table*}

\section{Spectra of strange hidden-charm and bottom tetra- and pentaquarks in the hadro-quarkonium model}
In this section, we discuss our results for the spectrum of heavy quarkonium-strange hadron bound states.

The binding energies are computed in the hadro-quarkonium model of Sec. \ref{Hadro-quarkonium Hamiltonian} and Refs. \cite{Anwar:2018bpu,Ferretti:2018kzy,Dubynskiy:2008mq} by solving the two-body eigenvalue problem of Eq. (\ref{eqn:Vhc}) via a finite differences algorithm \cite[Vol. 3, Sec. 16-6]{Feynman-Lectures}. As a check, the same results are also obtained by means of a numerical code based on the Multhopp method; see \cite[Sec. 2.4]{Richard}.
The values of the heavy quarkonium chromo-electric polarizabilities and light hadron radii used here are given in Table \ref{tab:Hq-Model-parameters}.

\subsection{Hidden-charm and hidden-bottom pentaquarks with strangeness in the hadro-quarkonium model}
\label{Hidden-charm pentaquarks with strangeness}
The first step of our investigation is the study of heavy quarkonium-hyperon bound states. 
Our findings are enlisted in Tables \ref{tab:hadro-quarkonium-spectrum-ccS} and \ref{tab:hadro-quarkonium-spectrum-ccX}.

It is worth noting that: I) According to our predictions, heavy-quarkonium-hyperon states may be deeply bound; II) In some cases, the $V_{\rm hq}$ potential well is deep enough to give rise to a heavy-quarkonium-baryon bound state and its radial excitation; III) Our results show a strong dependence on the hyperon's radius, $R_{\mathcal B}$. See Eq. (\ref{eqn:Vhc}). The theoretical predictions for $R_{\mathcal B}$'s are highly model dependent and span a relatively wide range \cite{Ledwig:2006gw,Kubis:1999xb,Buchmann:2002et,Carrillo-Serrano:2016igi}. 
However, the use of different values of the hyperon's radius does not change our first conclusion qualitatively. 
As an example, we consider the $\eta_{\rm c}(2S) \otimes \Sigma$  state.
If we extract the value of the $\Sigma$ radius from Ref. \cite{Carrillo-Serrano:2016igi}, $R_{\Sigma} = \frac{1}{2} \left(R_{\Sigma^+} + R_{\Sigma^-}\right) = 0.91$ fm, we get a binding energy $B_{\eta_{\rm c}(2S) \otimes \Sigma} = -294$ MeV; if we use the experimental value \cite{Tanabashi:2018oca}, $R_{\Sigma} = R_{\Sigma^-} = 0.780$ fm, we obtain $B_{\eta_{\rm c}(2S) \otimes \Sigma} = -492$ MeV. The previous results can be compared to our prediction from Table \ref{tab:hadro-quarkonium-spectrum-ccS}, $B_{\eta_{\rm c}(2S) \otimes \Sigma} = -355$ MeV, calculated with $R_{\Sigma} = \frac{1}{2} \left(R_{\Sigma^+} + R_{\Sigma^-}\right) = 0.863$ fm \cite{Ledwig:2006gw}; IV) In bottomonium-hyperon configurations, the presence of a heavier (nonrelativistic) $b \bar b$ pair is expected to make the hadro-bottomonium system more stable than the hadro-charmonium one due to kinetic energy suppression. This is why the strange hidden-bottom pentaquarks are more tightly bound than their hidden-charm counterparts; V) If we consider the second set of values for the bottomonium chromo-electric polarizabilities of Table \ref{tab:Hq-Model-parameters}, we get bottomonium-$\Sigma$ bound states characterized by very large binding energies. The hadro-quarkonium picture may break down in these specific cases. Thus, one may have to consider the possibility of a mixing between hadro-quarkonium and compact five-quark components:
\begin{equation}
	\label{eqn:H-mixing}
	H = \left( \begin{array}{cc} H_{\rm hq} & V_{\rm mixing} \\ V_{\rm mixing} & H_{\rm compact} \end{array} \right)
	\mbox{ }.
\end{equation}
Here, $H_{\rm hq} = V_{\rm hq} + T_{\rm hq}$ is the hadro-quarkonium Hamiltonian, with $T_{\rm hq}$ being the $\psi \mathcal H$ relative kinetic energy and $V_{\rm hq}$ the potential of Eq. (\ref{eqn:Vhc}); $H_{\rm compact}$ is an effective Hamiltonian, which describes a compact five-quark system; $V_{\rm mixing}$ is an off-diagonal interaction, which mixes hadro-quarkonium and compact five-quark components. 
\begin{table*}[htbp]
\begin{ruledtabular}
\begin{tabular}{cccccccccc}
Composition & & Quark content & & $\alpha_{\psi \psi}(n\ell)$ [GeV$^{-3}$] & & $J_{\rm tot}^P$ & & Mass (Binding) [MeV] \\
\hline
$\chi_{\rm c0}(1P) \otimes K$ & & $n \bar sc \bar c$ ($s \bar nc \bar c$) & & 11 & & $0^-$  & & 3886 ($-22$)  \\
$\eta_{\rm c}(2S) \otimes K$  & & $n \bar sc \bar c$ ($s \bar nc \bar c$)  & & 18 & & $0^+$ & & 3948 ($-183$)  \\
$\chi_{\rm c1}(1P) \otimes K$ & & $n \bar sc \bar c$ ($s \bar nc \bar c$) & & 11  & & $1^-$  & & 3981 ($-23$)  \\
$\psi(2S) \otimes K$               & & $n \bar sc \bar c$ ($s \bar nc \bar c$)  & & 18 & & $1^+$ & & 3996 ($-184$)  \\
$h_{\rm c}(1P) \otimes K$      & & $n \bar sc \bar c$ ($s \bar nc \bar c$) & & 11 & & $1^-$   & & 3996 ($-23$)  \\
$\chi_{\rm c2}(1P) \otimes K$ & & $n \bar sc \bar c$ ($s \bar nc \bar c$) & & 11 & & $2^-$   & & 4027 ($-23$)  \\
\hline
$\chi_{\rm c0}(1P) \otimes K^*$ & & $n \bar sc \bar c$ ($s \bar nc \bar c$) & & 11 & & $1^-$  & & 4155 ($-151$)  \\
$\eta_{\rm c}(2S) \otimes K^*$ & & $n \bar sc \bar c$ ($s \bar nc \bar c$)  & & 18 & & $1^+$ & & 4159 ($-370$)  \\
$\psi(2S) \otimes K^*$              & & $n \bar sc \bar c$ ($s \bar nc \bar c$)  & & 18 & & $0^+,1^+,2^+$ & & 4207 ($-371$)  \\
$\chi_{\rm c1}(1P) \otimes K^*$ & & $n \bar sc \bar c$ ($s \bar nc \bar c$) & & 11  & & $0^-,1^-,2^-$  & & 4250 ($-152$)  \\
$h_{\rm c}(1P) \otimes K^*$      & & $n \bar sc \bar c$ ($s \bar nc \bar c$) & & 11 & & $0^-,1^-,2^-$   & & 4265 ($-153$)  \\
$\chi_{\rm c2}(1P) \otimes K^*$ & & $n \bar sc \bar c$ ($s \bar nc \bar c$) & & 11 & & $1^-,2^-,3^-$   & & 4295 ($-153$)  \\
$\left[\eta_{\rm c}(2S) \otimes K^*\right]_{\ell_{\rm hq} = 1}$ & & $n \bar sc \bar c$ ($s \bar nc \bar c$) & & 18 & & $0^-,1^-,2^-$  & & 4436 ($-93$)   \\   
$\left[\psi(2S) \otimes K^*\right]_{\ell_{\rm hq} = 1}$ & & $n \bar sc \bar c$ ($s \bar nc \bar c$) & &18 & & $0^-,1^-,2^-,3^-$  & & 4484 ($-94$)   \\ 
\end{tabular}
\end{ruledtabular}
\caption{As Table \ref{tab:hadro-quarkonium-spectrum-ccS}, but for charmonium-$K$ and $K^*(892)$ bound states.}
\label{tab:hadro-quarkonium-spectrum-ccK}
\end{table*}

\begin{table*}[htbp]
\begin{ruledtabular}
\begin{tabular}{cccccccccc}
Composition & & Quark content & & $\alpha_{\psi \psi}(n\ell)$ [GeV$^{-3}$] & & $J_{\rm tot}^P$ & & Mass (Binding) [MeV] \\
\hline
$\eta_{\rm b}(2S) \otimes K$ & & $n \bar sb \bar b$ ($s \bar nb \bar b$) & & 23 & & $0^+$ & & 10121 ($-372$)   \\
$\Upsilon(2S) \otimes K$ & & $n \bar sb \bar b$ ($s \bar nb \bar b$) & & 23 & & $1^+$ & & 10145 ($-372$)   \\
$\chi_{\rm b0}(1P) \otimes K$ & & $n \bar sb \bar b$ ($s \bar nb \bar b$) & & 14 & & $0^-$ & & 10254 ($-99$)   \\
$\chi_{\rm b1}(1P) \otimes K$ & & $n \bar sb \bar b$ ($s \bar nb \bar b$) & & 14 & & $1^-$ & & 10288 ($-99$)   \\
$h_{\rm b}(1P) \otimes K$ & & $n \bar sb \bar b$ ($s \bar nb \bar b$) & & 14 & & $1^-$ & & 10294 ($-99$)  \\
$\chi_{\rm b2}(1P) \otimes K$ & & $n \bar sb \bar b$ ($s \bar nb \bar b$) & & 14 & & $2^-$ & & 10307 ($-99$)   \\
\hline
$\eta_{\rm b}(2S) \otimes K$ & & $n \bar sb \bar b$ ($s \bar nb \bar b$) & & 33 & & $0^+$ & & 9755 ($-738$)   \\
$\Upsilon(2S) \otimes K$ & & $n \bar sb \bar b$ ($s \bar nb \bar b$) & & 33 & & $1^+$ & & 9779 ($-738$)   \\
$\chi_{\rm b0}(1P) \otimes K$ & & $n \bar sb \bar b$ ($s \bar nb \bar b$) & & 21 & & $0^-$ & & 10049 ($-305$)   \\
$\chi_{\rm b1}(1P) \otimes K$ & & $n \bar sb \bar b$ ($s \bar nb \bar b$) & & 21 & & $1^-$ & & 10082 ($-305$)   \\
$h_{\rm b}(1P) \otimes K$ & & $n \bar sb \bar b$ ($s \bar nb \bar b$) & & 21 & & $1^-$ & & 10088 ($-305$)  \\
$\chi_{\rm b2}(1P) \otimes K$ & & $n \bar sb \bar b$ ($s \bar nb \bar b$) & & 21 & & $2^-$ & & 10101 ($-305$)   \\
$\left[\eta_{\rm b}(2S) \otimes K\right]_{\ell_{\rm hq} = 1}$ & & $n \bar sb \bar b$ ($s \bar nb \bar b$) & & 33 & & $1^-$  & & 10425 ($-68$)   \\                                              
$\left[\Upsilon(2S) \otimes K\right]_{\ell_{\rm hq} = 1}$ & & $n \bar sb \bar b$ ($s \bar nb \bar b$) & & 33 & & $0^-$, $1^-$ or $2^-$ & & 10449 ($-68$)  \\
\hline
$\eta_{\rm b}(2S) \otimes K^*$ & & $n \bar sb \bar b$ ($s \bar nb \bar b$) & & 23 & & $1^+$ & & 10323 ($-568$)   \\
$\Upsilon(2S) \otimes K^*$ & & $n \bar sb \bar b$ ($s \bar nb \bar b$) & & 23 & & $0^+,1^+,2^+$ & & 10347 ($-568$)   \\
$\chi_{\rm b0}(1P) \otimes K^*$ & & $n \bar sb \bar b$ ($s \bar nb \bar b$) & & 14 & & $1^-$ & & 10484 ($-267$)   \\
$\chi_{\rm b1}(1P) \otimes K^*$ & & $n \bar sb \bar b$ ($s \bar nb \bar b$) & & 14 & & $0^-,1^-,2^-$ & & 10517 ($-267$)   \\
$h_{\rm b}(1P) \otimes K^*$ & & $n \bar sb \bar b$ ($s \bar nb \bar b$) & & 14 & & $0^-,1^-,2^-$ & & 10524 ($-267$)  \\
$\chi_{\rm b2}(1P) \otimes K^*$ & & $n \bar sb \bar b$ ($s \bar nb \bar b$) & & 14 & & $1^-,2^-,3^-$ & & 10536 ($-267$)   \\
$\left[\eta_{\rm b}(2S) \otimes K^*\right]_{\ell_{\rm hq} = 1}$ & & $n \bar sb \bar b$ ($s \bar nb \bar b$) & & 23 & & $0^-,1^-,2^-$  & & 10604 ($-286$)   \\
$\left[\Upsilon(2S) \otimes K^*\right]_{\ell_{\rm hq} = 1}$ & & $n \bar sb \bar b$ ($s \bar nb \bar b$) & & 23 & & $0^-,1^-,2^-,3^-$  & & 10629 ($-286$)   \\   
$\left[\chi_{\rm b0}(1P) \otimes K^*\right]_{\ell_{\rm hq} = 1}$ & & $n \bar sb \bar b$ ($s \bar nb \bar b$) & & 14 & & $0^+,1^+,2^+$ & & 10711 ($-40$)   \\
$\left[\chi_{\rm b1}(1P) \otimes K^*\right]_{\ell_{\rm hq} = 1}$ & & $n \bar sb \bar b$ ($s \bar nb \bar b$) & & 14 & & $0^+,1^+,2^+,3^+$ & & 10745 ($-40$)   \\
$\left[h_{\rm b}(1P) \otimes K^*\right]_{\ell_{\rm hq} = 1}$ & & $n \bar sb \bar b$ ($s \bar nb \bar b$) & & 14 & & $0^+,1^+,2^+,3^+$ & & 10751 ($-40$)  \\ 
$\left[\chi_{\rm b2}(1P) \otimes K^*\right]_{\ell_{\rm hq} = 1}$ & & $n \bar sb \bar b$ ($s \bar nb \bar b$) & & 14 & & $0^+,1^+,2^+,3^+,4^+$ & & 10764 ($-40$)   \\
\hline
$\eta_{\rm b}(2S) \otimes K^*$ & & $n \bar sb \bar b$ ($s \bar nb \bar b$) & & 33 & & $1^+$ & & 9974 ($-917$); 10786 ($-105$)   \\
$\Upsilon(2S) \otimes K^*$      & & $n \bar sb \bar b$ ($s \bar nb \bar b$) & & 33 & & $0^+,1^+,2^+$ & & 9998 ($-917$); 10810 ($-105$)   \\
$\chi_{\rm b0}(1P) \otimes K^*$ & & $n \bar sb \bar b$ ($s \bar nb \bar b$) & & 21 & & $1^-$ & & 10252 ($-499$)   \\                                                                                                                                                  
$\left[\eta_{\rm b}(2S) \otimes K^*\right]_{\ell_{\rm hq} = 1}$ & & $n \bar sb \bar b$ ($s \bar nb \bar b$) & & 33 & & $0^-,1^-,2^-$  & & 10284 ($-607$)   \\
$\chi_{\rm b1}(1P) \otimes K^*$ & & $n \bar sb \bar b$ ($s \bar nb \bar b$) & & 21 & & $0^-,1^-,2^-$ & & 10285 ($-499$)   \\ 
$h_{\rm b}(1P) \otimes K^*$ & & $n \bar sb \bar b$ ($s \bar nb \bar b$) & & 21 & & $0^-,1^-,2^-$ & & 10292 ($-499$)  \\
$\chi_{\rm b2}(1P) \otimes K^*$ & & $n \bar sb \bar b$ ($s \bar nb \bar b$) & & 21 & & $1^-,2^-,3^-$ & & 10305 ($-499$)   \\
$\left[\Upsilon(2S) \otimes K^*\right]_{\ell_{\rm hq} = 1}$ & & $n \bar sb \bar b$ ($s \bar nb \bar b$) & & 33 & & $0^-,1^-,2^-,3^-$  & & 10308 ($-607$)   \\ 
$\left[\chi_{\rm b0}(1P) \otimes K^*\right]_{\ell_{\rm hq} = 1}$ & & $n \bar sb \bar b$ ($s \bar nb \bar b$) & & 21 & & $0^+,1^+,2^+$ & & 10525 ($-226$)   \\
$\left[\chi_{\rm b1}(1P) \otimes K^*\right]_{\ell_{\rm hq} = 1}$ & & $n \bar sb \bar b$ ($s \bar nb \bar b$) & & 21 & & $0^+,1^+,2^+,3^+$ & & 10558 ($-226$)   \\
$\left[h_{\rm b}(1P) \otimes K^*\right]_{\ell_{\rm hq} = 1}$ & & $n \bar sb \bar b$ ($s \bar nb \bar b$) & & 21 & & $0^+,1^+,2^+,3^+$ & & 10565 ($-226$)  \\  
$\left[\chi_{\rm b2}(1P) \otimes K^*\right]_{\ell_{\rm hq} = 1}$ & & $n \bar sb \bar b$ ($s \bar nb \bar b$) & & 21 & & $0^+,1^+,2^+,3^+,4^+$ & & 10578 ($-226$)   \\                                                
\end{tabular}
\end{ruledtabular}
\caption{As Table \ref{tab:hadro-quarkonium-spectrum-ccS}, but for bottomonium-$K$ and $K^*(892)$ bound states.}
\label{tab:hadro-quarkonium-spectrum-ccK*}
\end{table*}

\subsection{Hidden-charm and hidden-bottom tetraquarks with strangeness in the hadro-quarkonium model}
\label{Hidden-charm and hidden-bottom tetraquarks with strangeness in the hadro-quarkonium model}
As a second step, we study heavy quarkonium-kaon and $K^*$ configurations.
Our findings are enlisted in Tables \ref{tab:hadro-quarkonium-spectrum-ccK} and \ref{tab:hadro-quarkonium-spectrum-ccK*}. 

Heavy quarkonium-kaon bound states show similar features as the heavy-light pentaquarks of Sec. \ref{Hidden-charm pentaquarks with strangeness}.
In particular, one can notice that: I) The hadro-quarkonium interaction, Eq. (\ref{eqn:Vhc}), may determine the emergence of deeply-bound charmonium-kaon tetraquark configurations; II) Even more stable configurations are the bottomonium-kaon ones; III) In both previous cases, if one substitutes the kaon with the $K^*$, one obtains extremely stable systems. As discussed in Sec. \ref{Hidden-charm pentaquarks with strangeness}, a more realistic description of $\psi K^*$ systems may be accomplished by making use of the Hamiltonian (\ref{eqn:H-mixing}), where one also takes mixing effects between hadro-quarkonium and compact tetraquark components into account. Compact heavy-light tetraquarks have been extensively studied. For example, see the potential model calculations of Refs. \cite{Ebert:2008wm,Lu:2016cwr,Anwar:2018sol,Bedolla:2019zwg,Yang:2019itm} and Secs. \ref{TQ-Model} and \ref{Spectra of strange hidden-charm and bottom tetraquarks in the compact tetraquark model}.

The quality of the approximation of neglecting mixing effects between the heavy, $\psi$, and the light, $\mathcal H$, hadron components in the hadro-charmonium states of Table \ref{tab:hadro-quarkonium-spectrum-ccK} can be evaluated by calculating the wave function overlap of the previous components at the hadro-quarkonium center 
\begin{equation}
	\label{eqn:P_overlap}
	P_{\rm overlap} = \int_0^{R_{\mathcal H}} d^{3}r \mbox{ } \Psi_\psi({\bf r}) \Psi_{\mathcal H}({\bf r}) \mbox{ }.
\end{equation}	 
Here, $R_{\mathcal H}$ is the light hadron's radius and $\Psi_\psi({\bf r})$ and $\Psi_{\mathcal H}({\bf r})$ are the wave functions of the heavy and light hadro-quarkonium constituents, respectively, extracted from the relativized QM \cite{Godfrey:1985xj}.
If we restrict to the case of kaon-charmonium bound states, the heavy and light hadro-quarkonium's constituents can only be $1P$ or $2S$ charmonia (heavy component) and $1S$ $K$ or $K^*$ mesons (light component); see Table IV. 

We consider two different examples, $\chi_{\rm c0}(1P) \otimes K$ and $\eta_{\rm c}(2S) \otimes K$. All the other combinations of heavy and light mesons are analogous to the previous ones, because we expect the radial wave functions of all the other $\chi_{\rm c}(1P)$ states to be very similar to that of the $\chi_{\rm c0}(1P)$, namely $\Psi_{\chi_{\rm c0}(1P)}(r) \simeq \Psi_{h_{\rm c}(1P)}(r) \simeq \Psi_{\chi_{\rm c1}(1P)}(r) \simeq \Psi_{\chi_{\rm c2}(1P)}(r)$; analogously, we expect that $\Psi_{\psi(2S)}(r) \simeq \Psi_{\eta_{\rm c}(2S)}(r)$ and $\Psi_{K}(r) \simeq \Psi_{K^*}(r)$.
By calculating the overlap integral of Eq. (\ref{eqn:P_overlap}), we get $P_{\rm overlap}[\chi_{\rm c0}(1P) \otimes K] = 0$ and $P_{\rm overlap}[\eta_{\rm c}(2S) \otimes K] = 0.01$.

In conclusion, the previous test would indicate that in $\chi_{\rm c0}(1P) \otimes K$ and $\eta_{\rm c}(2S) \otimes K$ bound states the approximations we considered are acceptable ones and that there should be no substantial mixing among the heavy and light components.

\section{Relativized Diquark Model}
\label{TQ-Model}
We describe tetraquarks as color-antitriplet ($\bar 3_c$) diquark and color-triplet ($3_c$) antidiquark ($\mathcal D \bar {\mathcal D}$) bound states.
We also assume the constituents, $\mathcal D$ and $\bar {\mathcal D}$, to be inert against internal spatial excitations \cite{Anselmino:1992vg,Santopinto:2004hw,Ferretti:2011zz,Santopinto:2014opa}.
Consequently, the internal dynamics of the $\mathcal D \bar {\mathcal D}$ system can be described by means of a single relative coordinate $\bf{r}_{\rm rel}$ with conjugate momentum ${\bf q}_{\rm rel}$.

The Hamiltonian of the system is given by \cite{Anwar:2017toa,Anwar:2018sol,Bedolla:2019zwg}
\begin{subequations}
\label{eqn:Hmodel}	
\begin{align}
   	\mathcal{H}^{\rm REL} & = T + V({\bf r}_{\rm rel})\,,\\
         T & = \sqrt{{\mathbf q}_{\rm rel}^2+m_{{\mathcal D}_a}^2} + \sqrt{{\mathbf q}_{\rm rel}^2+m_{\bar {\mathcal D}_b}^2},
\end{align}
\end{subequations}
where the potential
\begin{widetext}
\begin{equation}
	\label{eqn:Vr12-new}  
	\begin{array}{rcl}
		V(r_{\rm rel}) & = & \beta r_{\rm rel} + G(r_{\rm rel}) +
    \frac{2 {\bf S}_{D_a} \cdot {\bf S}_{\bar D_b}}{3 m_{{\mathcal D}_a} m_{{\bar {\mathcal D}}_b}}
		\mbox{ } \nabla^2 G(r_{\rm rel}) 
		- \frac{1}{3 m_{{\mathcal D}_a} m_{\bar {\mathcal D}_b}} \left(3 {\bf S}_{{\mathcal D}_a} \cdot \hat r_{\rm rel}
		 \mbox{ } {\bf S}_{\bar {\mathcal D}_b} \cdot \hat r_{\rm rel} - {\bf S}_{{\mathcal D}_a} \cdot 
		 {\bf S}_{\bar {\mathcal D}_b}\right) \\
		& \times & \left(\frac{\partial^2}{\partial r_{\rm rel}^2}
		- \frac{1}{r_{\rm rel}} \frac{\partial}{\partial r_{\rm rel}}\right) G(r_{\rm rel}) + \Delta E \mbox{ },
	\end{array}
\end{equation}
is the sum of linear-confinement and one-gluon exchange (OGE) terms \cite{Celmaster:1977vh,Godfrey:1985xj,Capstick:1986bm, Anwar:2018sol}.
The Coulomb-like part is \cite{Godfrey:1985xj,Capstick:1986bm}
\begin{equation}
	\label{eqn:G(r)}
	G(r_{\rm rel}) = - \frac{4 \alpha_{\rm s}(r_{\rm rel})}{3 r_{\rm rel}} =
	- \sum_k \frac{4 \alpha_k}{3 r_{\rm rel}} \mbox{ Erf}(\tau_{ {\mathcal D}_a \bar {\mathcal D}_b k} \,r_{\rm rel})  \mbox{ },
\end{equation}
where Erf is the error function and \cite{Godfrey:1985xj,Capstick:1986bm}
\begin{equation}
	\label{eqn:sigma-DD}
	\tau_{ {\mathcal D}_a \bar {\mathcal D}_b\, k} = \frac{\gamma_k \sigma_{{\mathcal D}_a \bar {\mathcal D}_b}}
	{\sqrt{\sigma_{{\mathcal D}_a \bar {\mathcal D}_b}^2+\gamma_k^2}} \mbox{ };
	\mbox{ } \mbox{ } \sigma_{{\mathcal D}_a \bar {\mathcal D}_b} = \sqrt{\frac{1}{2} \sigma_0^2 \left[1 
	+ \left(\frac{4 m_{{\mathcal D}_a} m_{\bar {\mathcal D}_b}}{(m_{{\mathcal D}_a}
	+ m_{\bar {\mathcal D}_b})^2}\right)^4\right] 
	+ s^2 \left(\frac{2 m_{{\mathcal D}_a} m_{\bar {\mathcal D}_b}}{m_{{\mathcal D}_a} 
	+ m_{\bar {\mathcal D}_b}}\right)^2}
	\mbox{ }.
\end{equation}
\end{widetext}
The model parameters are listed in Table~\ref{tab:Model-parameters}.
The strength of the linear confining interaction, $\beta$, and the value of the constant, $\Delta E$, in Eq.\,\eqref{eqn:Vr12-new} are taken from \cite[Table I]{Anwar:2018sol}; the values of the parameters $\alpha_k$ and $\gamma_k$ ($k = 1,2,3$), $\sigma_0$ and $s$ are extracted from Refs. \cite{Godfrey:1985xj, Capstick:1986bm}.
The masses of the scalar and axial-vector diquarks $cn$, $cs$, $bn$ and $bs$, are taken from Refs. \cite{Anwar:2018sol,Bedolla:2019zwg,Ferretti:2019zyh}.

Therefore, the results we report below are parameter-free predictions.
The present model was previously used to calculate the spectrum of hidden-charm \cite{Anwar:2018sol} and fully-heavy tetraquarks \cite{Anwar:2017toa,Bedolla:2019zwg}.
\begin{table}[htbp]
\begin{ruledtabular}
\begin{tabular}{cccc}
Parameter  & Value   & Parameter  & Value \\
\hline
$\alpha_1$ & 0.25                   & $\gamma_1$ & 2.53 fm$^{-1}$  \\
$\alpha_2$ & 0.15                   & $\gamma_2$ & 8.01 fm$^{-1}$  \\
$\alpha_3$ & 0.20                   & $\gamma_3$ & 80.1 fm$^{-1}$  \\
$\sigma_0$ & 9.29 fm$^{-1}$  & $s$ & 1.55  \\
$\beta$        & 3.90 fm$^{-2}$  & $\Delta E$    & $-370$ MeV  \\
$M_{cn}^{\rm sc}$ & 1933 MeV & $M_{cn}^{\rm av}$ $\ $& 2250 MeV  \\
$M_{cs}^{\rm sc}$ & 2229 MeV & $M_{cs}^{\rm av}$ $\ $& 2264 MeV  \\
$M_{bn}^{\rm sc}$ & 5451 MeV & $M_{bn}^{\rm av}$ $\ $& 5465 MeV  \\
$M_{bs}^{\rm sc}$ & 5572 MeV & $M_{bs}^{\rm av}$ $\ $& 5585 MeV  \\
\end{tabular}
\end{ruledtabular}
\caption{Parameters of the Hamiltonian \eqref{eqn:Hmodel}. Here $n = u$ or $d$ and the superscripts ``sc'' and ``av'' indicate scalar and axial-vector diquarks, respectively.}
\label{tab:Model-parameters}
\end{table}

\begin{table*}[htbp]
\begin{ruledtabular}
\begin{center}
\begin{small}
\begin{tabular}{|ccc||ccc||ccc|}
\multicolumn{9}{|c|}{$\bf cs \bar c \bar n$ }\\
$J^{PC}$   & $N[(S_D,S_{\bar D})S,L]J$ & $E^{\rm th}$ [MeV] & $J^{PC}$   & $N[(S_D,S_{\bar D})S,L]J$ & $E^{\rm th}$ [MeV] & $J^{PC}$   & $N[(S_D,S_{\bar D})S,L]J$ & $E^{\rm th}$ [MeV] \\
\hline
$0^{++}$ &  $1[(1,1)0,0]0$ & 3657 & $1^{++}$ & $1[(1,0)1,0]1$ & 4016 & $0^{-+}$ &  $1[(1,0)1,1]0$ & 4396 \\
$0^{++}$ &  $1[(0,0)0,0]0$ & 3852 & $1^{++}$ & $2[(1,0)1,0]1$ & 4544 & $0^{-+}$ &  $1[(1,1)1,1]0$ & 4580 \\
$0^{++}$ &  $2[(0,0)0,0]0$ & 4383 & $1^{++}$ & $1[(1,0)1,2]1$ & 4658 & $0^{-+}$ &  $2[(1,0)1,1]0$ & 4783 \\
$0^{++}$ &  $2[(1,1)0,0]0$ & 4496 & $1^{++}$ & $1[(1,1)2,2]1$ & 4825 & $0^{-+}$ &  $2[(1,1)1,1]0$ & 4960 \\
$0^{++}$ &  $3[(0,0)0,0]0$ & 4747 & $1^{++}$ & $3[(1,0)1,0]1$ & 4906 & $0^{-+}$ &  $3[(1,0)1,1]0$ & 5093 \\
$0^{++}$ &  $1[(1,1)2,2]0$ & 4830 & $1^{++}$ & $2[(1,0)1,2]1$ & 4982 & $0^{-+}$ &  $3[(1,1)1,1]0$ & 5265 \\
$0^{++}$ &  $3[(1,1)0,0]0$ & 4913 & $1^{++}$ & $2[(1,1)2,2]1$ & 5147 & & & \\
$0^{++}$ &  $2[(1,1)2,2]0$ & 5151 & $1^{++}$ & $3[(1,0)1,2]1$ & 5261 & & & \\
$0^{++}$ &  $3[(1,1)2,2]0$ & 5427 & $1^{++}$ & $3[(1,1)2,2]1$ & 5423 & & & \\
\hline
$1^{--}$  & $1[(0,0)0,1]1$  & 4234 & $2^{++}$ & $1[(1,1)2,0]2$ & 4232 & $1^{+-}$ & $1[(1,0)1,0]1$ & 4016 \\
$1^{--}$  & $1[(1,0)1,1]1$  & 4396 & $2^{++}$ & $1[(0,0)0,2]2$ & 4497 & $1^{+-}$ & $1[(1,1)1,0]1$ & 4061 \\
$1^{--}$  & $1[(1,1)0,1]1$  & 4558 & $2^{++}$ & $2[(1,1)2,0]2$ & 4739 & $1^{+-}$ & $2[(1,0)1,0]1$ & 4544 \\
$1^{--}$  & $1[(1,1)2,1]1$  & 4583 & $2^{++}$ & $1[(1,1)2,2]2$ & 4818 & $1^{+-}$ & $2[(1,1)1,0]1$ & 4637 \\
$1^{--}$  & $2[(0,0)0,1]1$  & 4622 & $2^{++}$ & $1[(1,1)0,2]2$ & 4819 & $1^{+-}$ & $1[(1,0)1,2]1$ & 4658 \\
$1^{--}$  & $2[(1,0)1,1]1$  & 4783 & $2^{++}$ & $2[(0,0)0,2]2$ & 4824 & $1^{+-}$ & $1[(1,1)1,2]1$ & 4822 \\
$1^{--}$  & $2[(1,1)0,1]1$  & 4942 & $2^{++}$ & $3[(1,1)2,0]2$ & 5092 & $1^{+-}$ & $3[(1,0)1,0]1$ & 4906 \\
$1^{--}$  & $2[(1,1)2,1]1$  & 4962 & $2^{++}$ & $3[(0,0)0,2]2$ & 5105 & $1^{+-}$ & $2[(1,0)1,2]1$ & 4982 \\
$1^{--}$  & $3[(0,0)0,1]1$  & 4935 & $2^{++}$ & $2[(1,1)0,2]2$ & 5140 & $1^{+-}$ & $3[(1,1)1,0]1$ & 5010 \\
$1^{--}$  & $3[(1,0)1,1]1$  & 5093 & $2^{++}$ & $2[(1,1)2,2]2$ & 5140 & $1^{+-}$ & $2[(1,1)1,2]1$ & 5144 \\
$1^{--}$  & $3[(1,1)0,1]1$  & 5250 & $2^{++}$ & $3[(1,1)0,2]2$ & 5416 & $1^{+-}$ & $3[(1,0)1,2]1$ & 5261 \\
$1^{--}$  & $3[(1,1)2,1]1$  & 5268 & $2^{++}$ & $3[(1,1)2,2]2$ & 5416 & $1^{+-}$ & $3[(1,1)1,2]1$ & 5420 \\
\hline
$0^{--}$ &  $1[(1,0)1,1]0$ & 4396 & & & & & & \\
$0^{--}$ &  $2[(1,0)1,1]0$ & 4783 & & & & & & \\
$0^{--}$ &  $3[(1,0)1,1]0$ & 5093 & & & & & & \\
\end{tabular}
\end{small}
\end{center}
\end{ruledtabular}
\caption{Masses of $cs \bar c \bar n$ ($cn \bar c \bar s$) tetraquarks, obtained by solving the eigenvalue problem of Eq. (\ref{eqn:Hmodel}). We report states up to the second radial excitation. They are labelled thus: $N$ is the radial quantum number; $S_{\rm D}$, $S_{\bar {\rm D}}$ are the spin of the diquark and antidiquark, respectively, coupled to the total spin of the meson, $S$; the latter is coupled to the orbital angular momentum, $L$, to get the total angular momentum of the tetraquark, $J$. In the case of  scalar-axial-vector diquark configurations, there are two possible ways of combining diquarks to get a tetraquark with strangeness $\mathcal S = \pm 1$. They are $[cn] \{\bar c \bar s\}$ and $[cs] \{\bar c \bar n\}$, where the diquarks in square brackets are scalar and those in curly brackets are axial-vector. In these cases, the values of the tetraquark masses shown are the average of the energies corresponding to the previous $[cn] \{\bar c \bar s\}$ and $[cs] \{\bar c \bar n\}$ spin-flavor configurations.}
\label{tab:cncs-spectrum}
\end{table*}

\section{Masses of $cs \bar c \bar n$ and $bs \bar b \bar n$ states in the compact tetraquark model}
\label{Spectra of strange hidden-charm and bottom tetraquarks in the compact tetraquark model}
Below, we provide results for the ground-state masses and the spectrum of strange hidden-charm ($cs \bar c \bar n$ and $cn \bar c \bar s$) and bottom ($bs \bar b \bar n$ and $bn \bar b \bar s$) tetraquarks in the compact tetraquark model of Refs. \cite{Anwar:2017toa,Anwar:2018sol,Bedolla:2019zwg} and Sec. \ref{TQ-Model}.

The tetraquark masses are obtained by solving the eigenvalue problem of Eq. (\ref{eqn:Hmodel}) by means of a numerical variational procedure, based on harmonic oscillator trial wave functions. 
This variational method was previously applied to meson and baryon spectroscopy \cite{Ferretti:2011zz,Santopinto:2014opa,Bedolla:2019zwg,Anwar:2017toa,Anwar:2018sol}.

\subsection{Ground-state energies of $cs \bar c \bar n$ and $bs \bar b \bar n$ tetraquarks}
Our starting point is the calculation of the ground-state masses of $cs \bar c \bar n$ ($cn \bar c \bar s$) and $bs \bar b \bar n$ ($bn \bar b \bar s$) tetraquark configurations.

In the first case, we obtain
\begin{equation}
	\label{eqn:Mass-cscn}
	M_{cs\bar c \bar n}^{\rm gs} = \left\{ \begin{array}{rl} 3.85 \mbox{ GeV } & (\mbox{sc-sc configuration}) \\
	3.66 \mbox{ GeV } & (\mbox{av-av configuration}) \end{array} \right.  \mbox{ },
\end{equation}
where the notations ``sc'' and ``av'' indicate scalar and axial-vector diquarks, respectively.
The previous values have be compared with the experimental energy of the $D \bar D_{\rm s}$ threshold, $3.84$ GeV \cite{Tanabashi:2018oca}.
It is interesting to observe that the av-av $cs\bar c \bar n$ tetraquark ground-state is around 200 MeV below the lowest energy hadro-charmonium $\eta_{\rm c} \otimes K$ state of Table \ref{tab:hadro-quarkonium-spectrum-ccK}, which lies at an energy of 3886 MeV.

In the second case, we get
\begin{equation}
	\label{eqn:Mass-bsbn}
	M_{bs\bar b \bar n}^{\rm gs} = \left\{ \begin{array}{rl} 10.41 \mbox{ GeV } & (\mbox{sc-sc configuration}) \\
	10.23  \mbox{ GeV } & (\mbox{av-av configuration}) \end{array} \right.  \mbox{ },
\end{equation}
to be compared with the $B \bar B_{\rm s}$ threshold energy, $10.65$ GeV \cite{Tanabashi:2018oca}.
Contrary to the $cs\bar c \bar n$ case, the av-av $bs\bar b \bar n$ tetraquark ground-state is above the lowest energy hadro-bottomonium $\eta_{\rm b} \otimes K$ state of Table \ref{tab:hadro-quarkonium-spectrum-ccK*}, which lies at an energy between 9.76 and 10.12 GeV depending on the input value of the chromo-electric polarizability.

According to the previous results, strange hidden-charm and bottom tetraquarks may be bound.
However, due to the largeness of the theoretical uncertainties on the $bs \bar b \bar n$ and, especially, $cs \bar c \bar n$ ground-state tetraquark masses, it is difficult to draw a definitive conclusion. 

\subsection{Spectra of $cs \bar c \bar n$ and $bs \bar b \bar n$ tetraquarks}
After discussing the possible emergence of $cs \bar c \bar n$ and $bs \bar b \bar n$ tetraquarks and their ground-state energies, the next step is to calculate the spectrum predicted by the Hamiltonian of Eq. (\ref{eqn:Hmodel}) with the model parameters of Table \ref{tab:Model-parameters}.
In Tables \ref{tab:cncs-spectrum} and \ref{tab:bnbs-spectrum} we report the masses of $cs \bar c \bar n$ ($cn \bar c \bar s$) and $bs \bar b \bar n$ ($bn \bar b \bar s$) compact tetraquarks, where $n = u$ or $d$, up to the second radial excitations.

As discussed in Ref. \cite{Bedolla:2019zwg}, these type of predictions may serve as benchmarks for other analyses with the goal of identifying model-dependent artifacts and develop a perspective on those predictions which might only be weakly sensitive to model details. Moreover, given the possibility that $J = 0^{++}$ tetraquarks may be more difficult to access experimentally than $1^{--}$ resonances, our predictions for $J \neq 0$ states may be useful in the experimental search for $cs \bar c \bar n$ and $bs \bar b \bar n$ tetraquark states.

The calculation of the spectrum is only the first step of a wider analysis, with the aim of understanding the possible formation and stability of compact tetraquark states. The following steps include the calculation of tetraquark decay amplitudes, production cross-sections, and the study of their production mechanisms.
When compared to the same observables calculated within other interpretations for $XYZ$ states (like the meson-meson molecular model, the hadro-quarkonium model and the UQM) and the experimental data, it will be possible to distinguish among the different interpretations and possibly rule out those which are not compatible with the experimental resuls.

In conclusion, even though the experimental search for strange hidden-charm and bottom tetraquarks may be challenging, the observation of these systems may be extremely useful to understand the quark structure of $XYZ$ exotic mesons.

\begin{table*}[htbp]
\begin{ruledtabular}
\begin{center}
\begin{small}
\begin{tabular}{|ccc||ccc||ccc|}
\multicolumn{9}{|c|}{$\bf bs\bar b \bar n$ }\\
$J^{PC}$   & $N[(S_D,S_{\bar D})S,L]J$ & $E^{\rm th}$ [MeV] & $J^{PC}$   & $N[(S_D,S_{\bar D})S,L]J$ & $E^{\rm th}$ [MeV] & $J^{PC}$   & $N[(S_D,S_{\bar D})S,L]J$ & $E^{\rm th}$ [MeV] \\\hline
$0^{++}$ &  $1[(1,1)0,0]0$ & 10234 & $1^{++}$ & $1[(1,0)1,0]1$ & 10420 & $0^{-+}$ &  $1[(1,0)1,1]0$ & 10804 \\
$0^{++}$ &  $1[(0,0)0,0]0$ & 10407 & $1^{++}$ & $2[(1,0)1,0]1$ & 10922 & $0^{-+}$ &  $1[(1,1)1,1]0$ & 10827 \\
$0^{++}$ &  $2[(1,1)0,0]0$ & 10848 & $1^{++}$ & $1[(1,0)1,2]1$ & 11039 & $0^{-+}$ &  $2[(1,0)1,1]0$ & 11139 \\
$0^{++}$ &  $2[(0,0)0,0]0$ & 10909 & $1^{++}$ & $1[(1,1)2,2]1$ & 11054 & $0^{-+}$ &  $2[(1,1)1,1]0$ & 11159 \\
$0^{++}$ &  $1[(1,1)2,2]0$ & 11056 & $1^{++}$ & $3[(1,0)1,0]1$ & 11235 & $0^{-+}$ &  $3[(1,0)1,1]0$ & 11398 \\
$0^{++}$ &  $3[(1,1)0,0]0$ & 11185 & $1^{++}$ & $2[(1,0)1,2]1$ & 11311 & $0^{-+}$ &  $3[(1,1)1,1]0$ & 11418 \\
$0^{++}$ &  $3[(0,0)0,0]0$ & 11222 & $1^{++}$ & $2[(1,1)2,2]1$ & 11326 & & & \\
$0^{++}$ &  $2[(1,1)2,2]0$ & 11328 & $1^{++}$ & $3[(1,0)1,2]1$ & 11539 & & & \\
$0^{++}$ &  $3[(1,1)2,2]0$ & 11556 & $1^{++}$ & $3[(1,1)2,2]1$ & 11554 & & & \\
\hline
$1^{--}$  & $1[(0,0)0,1]1$  & 10790 & $2^{++}$ & $1[(1,1)2,0]2$ & 10467 & $1^{+-}$ & $1[(1,1)1,0]1$ & 10373 \\
$1^{--}$  & $1[(1,0)1,1]1$  & 10804 & $2^{++}$ & $2[(1,1)2,0]2$ & 10952 & $1^{+-}$ & $1[(1,0)1,0]1$ & 10420 \\
$1^{--}$  & $1[(1,1)0,1]1$  & 10816 & $2^{++}$ & $1[(0,0)0,2]2$ & 11025 & $1^{+-}$ & $2[(1,1)1,0]1$ & 10906 \\
$1^{--}$  & $1[(1,1)2,1]1$  & 10828 & $2^{++}$ & $1[(1,1)0,2]2$ & 11051 & $1^{+-}$ & $2[(1,0)1,0]1$ & 10922 \\
$1^{--}$  & $2[(0,0)0,1]1$  & 11125 & $2^{++}$ & $1[(1,1)2,2]2$ & 11051 & $1^{+-}$ & $1[(1,0)1,2]1$ & 11039 \\
$1^{--}$  & $2[(1,0)1,1]1$  & 11139 & $2^{++}$ & $3[(1,1)2,0]2$ & 11261 & $1^{+-}$ & $1[(1,1)1,2]1$ & 11053 \\
$1^{--}$  & $2[(1,1)0,1]1$  & 11151 & $2^{++}$ & $2[(0,0)0,2]2$ & 11297 & $1^{+-}$ & $3[(1,1)1,0]1$ & 11226 \\
$1^{--}$  & $2[(1,1)2,1]1$  & 11160 & $2^{++}$ & $2[(1,1)0,2]2$ & 11323 & $1^{+-}$ & $3[(1,0)1,0]1$ & 11235 \\
$1^{--}$  & $3[(0,0)0,1]1$  & 11385 & $2^{++}$ & $2[(1,1)2,2]2$ & 11323 & $1^{+-}$ & $2[(1,0)1,2]1$ & 11311 \\
$1^{--}$  & $3[(1,0)1,1]1$  & 11398 & $2^{++}$ & $3[(0,0)0,2]2$ & 11526 & $1^{+-}$ & $2[(1,1)1,2]1$ & 11325 \\
$1^{--}$  & $3[(1,1)0,1]1$  & 11411 & $2^{++}$ & $3[(1,1)0,2]2$ & 11552 & $1^{+-}$ & $3[(1,0)1,2]1$ & 11539 \\
$1^{--}$  & $3[(1,1)2,1]1$  & 11418 & $2^{++}$ & $3[(1,1)2,2]2$ & 11552 & $1^{+-}$ & $3[(1,1)1,2]1$ & 11553 \\
\hline
$0^{--}$ &  $1[(1,0)1,1]0$ & 10804 & & & & & & \\
$0^{--}$ &  $2[(1,0)1,1]0$ & 11139 & & & & & & \\
$0^{--}$ &  $3[(1,0)1,1]0$ & 11398 & & & & & & \\
\end{tabular}
\end{small}
\end{center}
\end{ruledtabular}
\caption{As Table \ref{tab:cncs-spectrum}, but for $bs\bar b \bar n$ ($bn\bar b \bar s$) states.}
\label{tab:bnbs-spectrum}
\end{table*}

\section{Conclusion}
We calculated the spectrum of strange hidden-charm and bottom tetraquarks both in the hadro-quarkonium model of Refs. \cite{Anwar:2018bpu,Ferretti:2018kzy,Dubynskiy:2008mq} and the compact tetraquark model of Refs. \cite{Anwar:2017toa,Anwar:2018sol,Bedolla:2019zwg}. We also computed that of hidden-charm and bottom pentaquarks in the hadro-quarkonium model.
In particular, we discussed the possible emergence of $\eta_{\rm b,c}(2S)$-, $\psi(2S)$-, $\Upsilon(2S)$-, and $\chi_{\rm b,c}(1P)$-hyperon/kaon bound states and the possible formation of $cs\bar c \bar n$ and $bs \bar b \bar n$ tetraquarks as diquark-antidiquark bound states. 

Our results suggest that: I) strange hadro-quarkonium systems may be strongly bound. On the other hand, if the heavy quarkonium- ($\psi$) light hadron ($\mathcal H$) binding energies become too large, the hadro-quarkonium picture may break down. As a consequence, the $\psi$ and $\mathcal H$ components may overlap, and a compact four/five-quark system could be realized rather than a $\psi$-$\mathcal H$ bound state; II) both $cs\bar c \bar n$ and $bs \bar b \bar n$ compact tetraquarks may be bound, even though $bs \bar b \bar n$ configurations are more likely to manifest; III) in the case of $cs\bar c \bar n$ configurations, the compact tetraquark ground-state is around 200 MeV below the lowest energy hadro-charmonium state, $\eta_{\rm c} \otimes K$. On the contrary, in the $bs\bar b \bar n$ case the compact tetraquark ground-state is above the lowest energy hadro-bottomonium configuration, $\eta_{\rm b} \otimes K$; IV) by combining the conclusions discussed at points I) and II), we suggest the experimentalists to look for strange tetra- and pentaquark configurations with hidden-bottom. They should be more stable than their hidden-charm counterparts due to kinetic energy suppression; thus, there is a higher probability of observing them.

Finally, as pointed out in Ref. \cite{Voloshin:2019}, the meson-meson molecular model cannot be used to describe heavy-light tetraquarks with non-null strangeness content. The reason is that one-pion-exchange cannot take place between strange and nonstrange heavy mesons, like $B$ and $B_{\rm s}$.
Hidden-charm and bottom mesons with strangeness are also forbidden in the context of the Unquenched Quark Model (UQM) formalism. Indeed, one cannot dress heavy quarkonium $Q \bar Q$ states with $Q \bar s - n \bar Q$ or $Q \bar n - s \bar Q$ higher Fock components (where $n = u$ or $d$) by creating a light $n \bar n$ or $s \bar s$ pair with vacuum quantum numbers.
Tetraquarks with non-null strangeness content can only take place either in the compact tetraquark or hadroquarkonium models.
Therefore, a possible way to discriminate between the compact tetraquark and hadro-quarkonium models on one side and the molecular model and UQM interpretations on the other is the experimental search for strange hidden-charm and bottom four-quark states.

Our predictions for $P_{\rm c}$ and $P_{\rm b}$ pentaquarks with non-null strangeness content may be soon be tested by LHCb.

\begin{acknowledgments}
This work was supported by the U.S. Department of Energy (Grant No. DE-FG-02-91ER-40608) and the Academy of Finland, Project No. 320062.
\end{acknowledgments}

\end{document}